\documentclass[twocolumn,floatfix,showpacs,amsmath,amssymb,pra]{revtex4}
\usepackage{times}
\usepackage{graphicx}
\usepackage{dcolumn}
\usepackage{bm}

\newcommand{\Tr}{{\rm Tr}}

\newcommand{\ket}[1]{\left | \, #1 \right\rangle}
\newcommand{\bra}[1]{\left \langle #1 \, \right |}

\newcommand{\la}{\langle}
\newcommand{\ra}{\rangle}
\newcommand{\no}{\nonumber\\}

\begin{document}

\title{Stochastic exclusion processes versus coherent transport}

\author{Kristan Temme$^1$}
\thanks{email:kristan.temme@univie.ac.at}
\author{Michael M. Wolf$^2$}
\author{Frank Verstraete$^1$}
\affiliation{$^1$Faculty of Physics, University of Vienna, Boltzmanngasse 5, A-1090 Vienna, Austria\\
		$^2$Niels Bohr Institute, 2100 Copenhagen, Denmark}
\date{\today}

\begin{abstract} 
\noindent
Stochastic exclusion processes play an integral role in the physics of non-equilibrium statistical 
mechanics. These models are Markovian processes, described by a classical master equation.
In this paper a quantum mechanical version of a stochastic hopping process in one dimension
is formulated in terms of a quantum master equation. This allows the investigation of coherent and 
stochastic evolution in the same formal framework. The focus lies on the non-equilibrium steady state.
Two stochastic model systems are considered, the totally asymmetric exclusion process and the 
fully symmetric exclusion process. The steady state transport properties of these models is compared 
to the case with additional coherent evolution, generated by the $XX$-Hamiltonian.   
\end{abstract}

\maketitle
\section{Introduction}
\label{sec:intro}

Stochastic exclusion processes have been studied in statistical mechanics for a long time
\cite{Ligget85,Hinrichsen01}. These are simplified one-dimensional hopping  models,
that allow the study of non-equilibrium phenomena in many-particle systems. 
The exclusion processes are modeled by a classical master 
equation that determines the time evolution of the probability distribution.
The steady state of the master equation exhibits interesting non-equilibrium behavior, such 
as the presence of a current, non-equilibrium phase transitions and entire phases with a diverging 
correlation length\cite{ASEP1,ASEP2}. 
The presence of currents, such as the current of particles, energy or momentum, 
is a common feature of non-equilibrium steady states and can have profound effects on the correlations 
present in the system \cite{Curr1,Curr2}. Non-equilibrium systems can develop long-range correlations in the presence of 
a high current.\\
The asymmetric exclusion process (ASEP) as well as the symmetric exclusion process (SEP) are prime examples for such
model systems\cite{ASEP1,SEP1}. These processes describe the hopping of hard-core particles in a one-dimensional 
chain, only driven by the inflow and the outflow of particles at the boundaries of the chain. Here one considers 
open boundary conditions, where particles are injected at the first site and are removed at the last site $N$ of the chain.
The steady state properties are entirely determined by the inflow and outflow.
The dynamics of the particles in the bulk are given by translationally invariant hopping rates, that either constrain the 
hopping of particles to take place in only one direction (ASEP) or allow for a hopping in both directions (SEP).
\\
Transport properties of open quantum mechanical systems, on the other hand, are subject to recent research activities.
A general interest is placed on how external noise, generated by the environment, 
affects the coherent transport in the system. It has been found, that the presence of noise in the quantum mechanical 
systems can actually aid the transport process of excitations through heterogeneous environments \cite{Plenio1, Rebentrost}, 
such as bio-molecules. An optimal ratio between coherent transport and dephasing noise can be found. 
The dynamics of open quantum systems are generally formulated in terms of a 
Markovian Lindblad master equation that describes the time evolution of the density matrix \cite{Master}.
\\
In this article we want to investigate the interplay between stochastic transport processes and coherent transport present 
in the same system. Here we consider only the steady-state properties of the system. 
To treat both processes on equal footing, we incorporate the classical hopping terms into the quantum master equation.
The stochastic hopping is modeled by appropriately chosen quantum jump operators. 
Such a construction has also been used to find quantum master equations that describe a quantization of 
kinetic Ising models \cite{Lew}. These models obey detailed balance and allow for an exact solution.
Considering hopping models in this more general quantum framework allows now for additional quantum transport,
so to speak, on top of the classical hopping evolution. We can choose an arbitary, particle number conserving Hamiltonian 
to mediate the coherent transport and investigate the effect this quantum pertubation has on the classical hopping process.
\\
Our article is organized as follows: First, in section \ref{sec:master}, 
we introduce the hopping model and formulate the problem as a quantum
master equation. We choose in different instances either the ASEP or the SEP as the underlying stochastic 
process. In the following section \ref{sec:ASEP}, the quantum analog of the ASEP is treated numerically 
in the framework of matrix product density operators (MPDO). The master equation for a chain of $N=40$ sites 
is evolved in time, until the steady state is reached. The current, the particle density, as well as the particle 
density-density correlations are computed. In section \ref{sec:SEP} the SEP is considererd. 
The two-point correlation functions of the SEP can be calculated exactly in the steady state, and the scaling 
of the current for larger lattice sizes is investigated. The conclusion are then drawn in the final section 
\ref{sec:Concl}. 


\section{Formulation of the quantum master equation}
\label{sec:master}

We study a system of hard-core particles in a one dimensional chain of length $N$, where 
each site $(1 \leq k \leq N)$ can be either occupied or empty. In the spin chain picture 
this corresponds to either spin up (occupied), or spin down (empty). At the boundary $k=1$,
we allow for an inflow with a rate $\alpha$ and at $k=N$ for an outflow of particles, 
given by a rate $\beta$. The particles at each site are allowed to hop stochastically 
to the left with a rate $\varphi_L$ and to the right with a rate $\varphi_R$. 
In this article we consider only two cases. First, the fully asymmetric case (ASEP), where the particles only hop
to the right, $\varphi_R = 1$ and $\varphi_L=0$. Second, the fully symmetric case (SEP) where 
both hopping rates are equal $\varphi_L = \varphi_R = 1$.

\begin{figure}[tb]
\begin{center}
\includegraphics[width=0.45\textwidth]{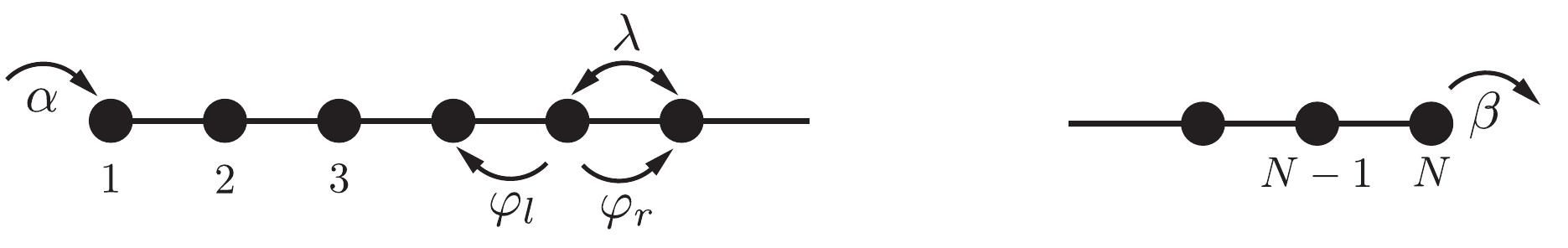}
\end{center}
\vspace*{-3ex}
\caption{spin-$1/2$ chain with jump-operators and coherent evolution. The inflow is given by rate $\alpha$, the outflow
with rate $\beta$, stochastic hopping to the left with rate $\varphi_L$ and to the right $\varphi_R$. The strength of the 
coherent evolution is given by $\lambda$}
\label{fig:EP_chain}
\end{figure}
These stochastic particle jumps correspond to the classical stochastic exclusion process
and are formulated in terms of quantum-jump operators $L_{\mu}$.  These operators govern the incoherent 
evolution of the quantum master equation. These jumps can be formulated in terms of spin-flip 
operations:
\begin{eqnarray}
L_{1}         &=& \sqrt{\alpha}\, \sigma^{-}\\ \nonumber
L_{N}         &=& \sqrt{\beta}\, \sigma^{+} \\ \nonumber
L^{R}_{k,k+1} &=& \sqrt{\varphi_R}\, \sigma^{-}_{k} \otimes \sigma^{+}_{k+1} \\ \nonumber
L^{L}_{k-1,k} &=& \sqrt{\varphi_L}\, \sigma^{+}_{k-1} \otimes \sigma^{-}_{k}.
\end{eqnarray}
Here, the $\sigma^{\pm}$ correspond to the Pauli raising and lowering operators. The master equation written with only 
these operators reproduces exactly the classical stochastic behavior with the appropriate jump rates. This can be seen,
when one restricts oneself to density matrices diagonal in the computational basis. With the chosen definition of the 
Lindblad operators, the classical master equation for the probabilities is reproduced.
 In this generalized framework, we can now also allow for an additional coherent evolution of the system, when choosing 
an appropriate Hamiltonian. The $XX$-Hamiltonian
\begin{equation}
 H = \lambda \sum_{k=1}^{N-1} \sigma_k^x \otimes\sigma_{k+1}^x +\sigma_k^y \otimes \sigma_{k+1}^y.
\end{equation}
gives rise to the free coherent evolution of the hardcore particles. Furthermore it has the property that it conserves the 
number of particles. The full quantum master equation for the density operator $\rho$ can be written as: 
\begin{eqnarray}
\label{master}
\partial_t \rho &=& -i\left [\rho , H \right] + \sum_{\mu}L_{\mu}\rho L_{\mu}^{\dagger}
 - \frac{1}{2}\left \{L_{\mu}^{\dagger}L_{\mu};\rho\right\} \\ \nonumber 
&\equiv & {\cal L}\left [\rho \right].
\end{eqnarray}
The central observables are the particle density $n_k = \sigma_k^{+}\sigma_k^{-}$ and 
the current $j_k$ of particles. To find the right expression for the particle current, we 
consider the continuity equation for the density $n_k$. The continuity equation is obtained 
in the Heisenberg picture when the adjoined of ${\cal L}$ is acting on $n_k$. 
\begin{eqnarray}
\partial_t n_k &=& {\cal L}^{\dagger}\left [n_k \right] \\ \nonumber
                     &=& -i\left [ H , n_k \right] + \sum_{\mu}L_{\mu}^{\dagger} n_k L_{\mu}
                             - \frac{1}{2}\left \{L_{\mu}^{\dagger}L_{\mu};n_k \right\}  
\end{eqnarray}  
Let us first consider only the (ASEP) with a single Lindblad operator in the bulk, since $\varphi_L=0$. 
We find, that $n_k$ obeys a continuity equation of the form:
\begin{eqnarray}
\partial_t n_k = \left (j^{co}_{k,k+1} + j^{st}_{k,k+1} \right) - \left(j^{co}_{k-1,k} + j^{st}_{k-1,k}\right).
\end{eqnarray}
We may now interpret the sum of the terms
\begin{eqnarray}
\label{curr}
 j^{co}_{k,k+1} &= \frac{2\lambda}{i}\left(\sigma^{-}_k\sigma^{+}_{k+1} - \sigma^{+}_k\sigma^{-}_{k+1}\right)\\ \nonumber
 j^{st}_{k,k+1} &= \varphi_R \left(\sigma^{+}_{k} \sigma^{-}_{k}\left(1-\sigma^{+}_{k+1}\sigma^{-}_{k+1}\right) \right),
\end{eqnarray}
as the total current-density $j_{k,k+1}$ of the system. Note, that there are two different contributions to the current, 
the coherent part $j^{co}$ due to the dynamics generated by the Hamiltonian  and the stochastic contribution 
$j^{st}$ originating from the quantum-jump operator induced hopping. We observe, that the stochastic contribution
to the current corresponds exactly to the current present in the classical model \cite{ASEP1,Derrida93}. 
\section{The Asymetric exclusion process}
\label{sec:ASEP}

\begin{figure}[tb]
\begin{center}
\includegraphics[width=0.30\textwidth]{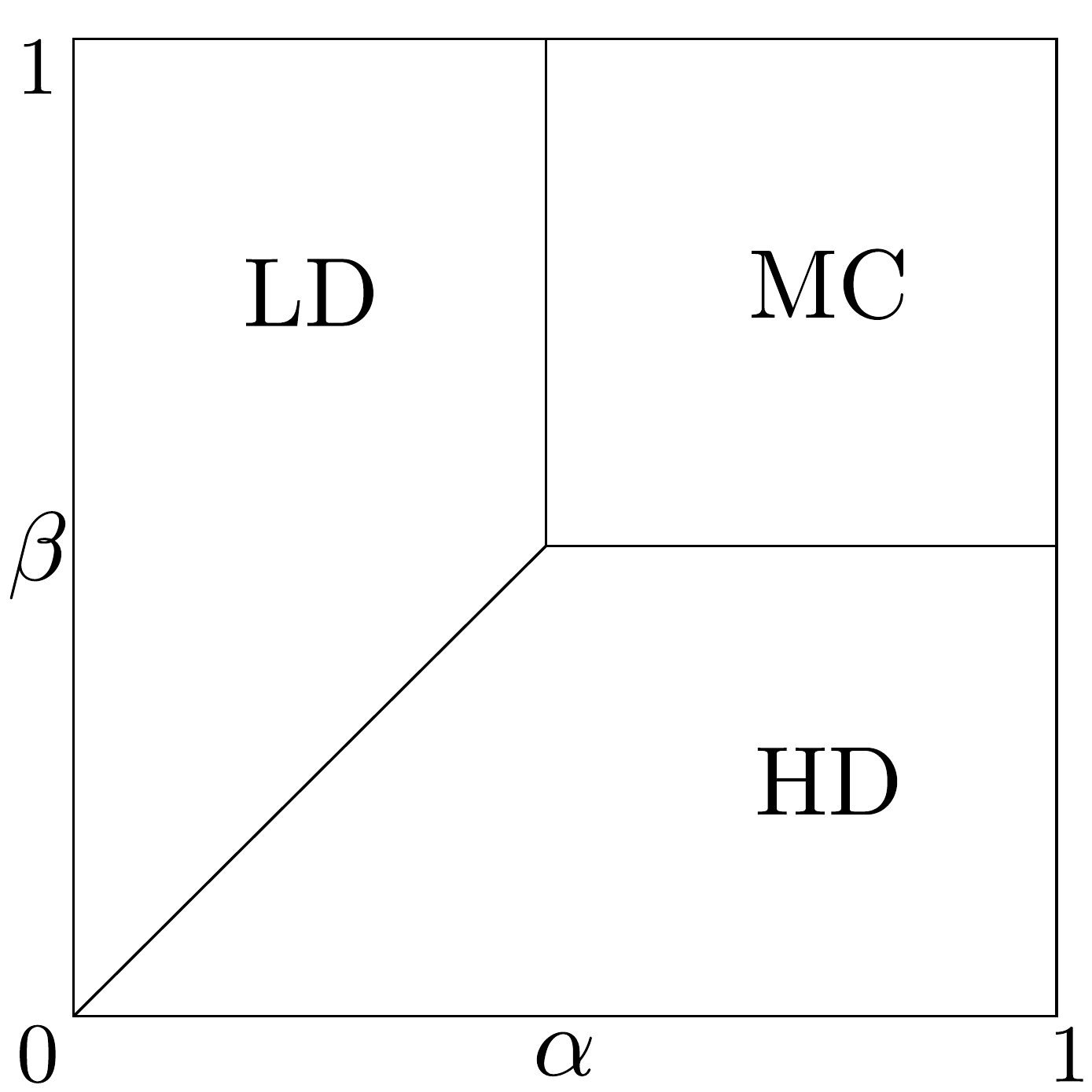}
\end{center}
\vspace*{-3ex}
\caption{Classical phase diagram of the asymmetric exclusion process. The ASEP processes three phases, LD (low-density),
HD (high-density) and the maximal current phase MC. }
\label{fig:phase_diagram}
\end{figure}

\begin{center}
\begin{figure}[ht]
\resizebox{\linewidth}{!}
{\includegraphics{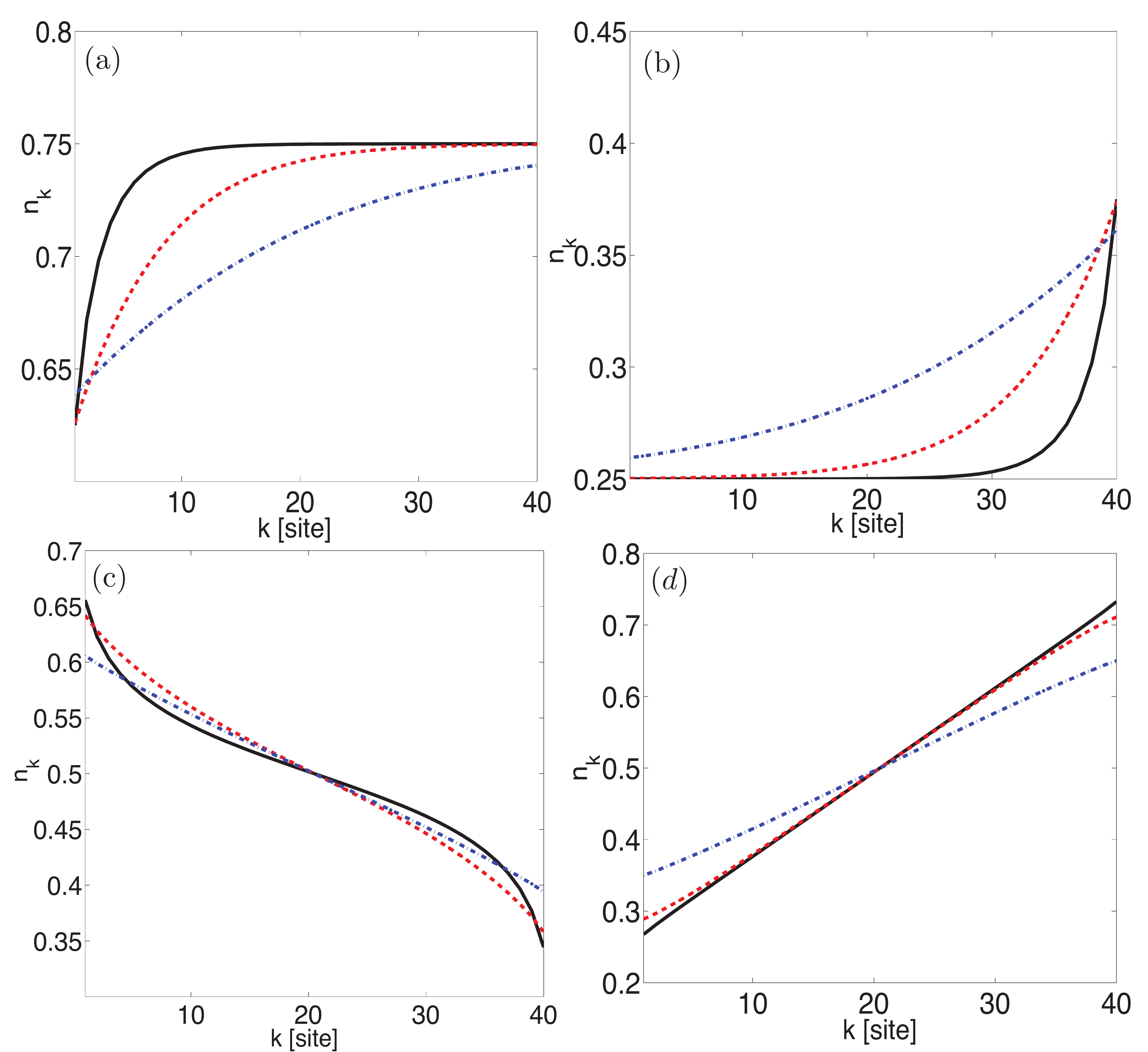}}
\caption{\label{fig:dens}(color online) 
Density distribution $\langle n_k \rangle$ for the different points $(\alpha,\beta)$, (a) HD $(1/2,1/4)$, (b) LD $(1/4,1/2)$ and
(c) MC $(3/4,3/4)$as well as (d) CL $(3/4,3/4)$. The black solid line corresponds to $\lambda = 0$, i.e. the classical solution.
The red dashed line corresponds to a quantum perturbation with $\lambda = 1/2$, and the blue dashed-dotted line to
a perturbation $\lambda = 1$.}
\end{figure}
\end{center}

\begin{center}
\begin{figure}[ht]
\resizebox{0.95\linewidth}{!}
{\includegraphics{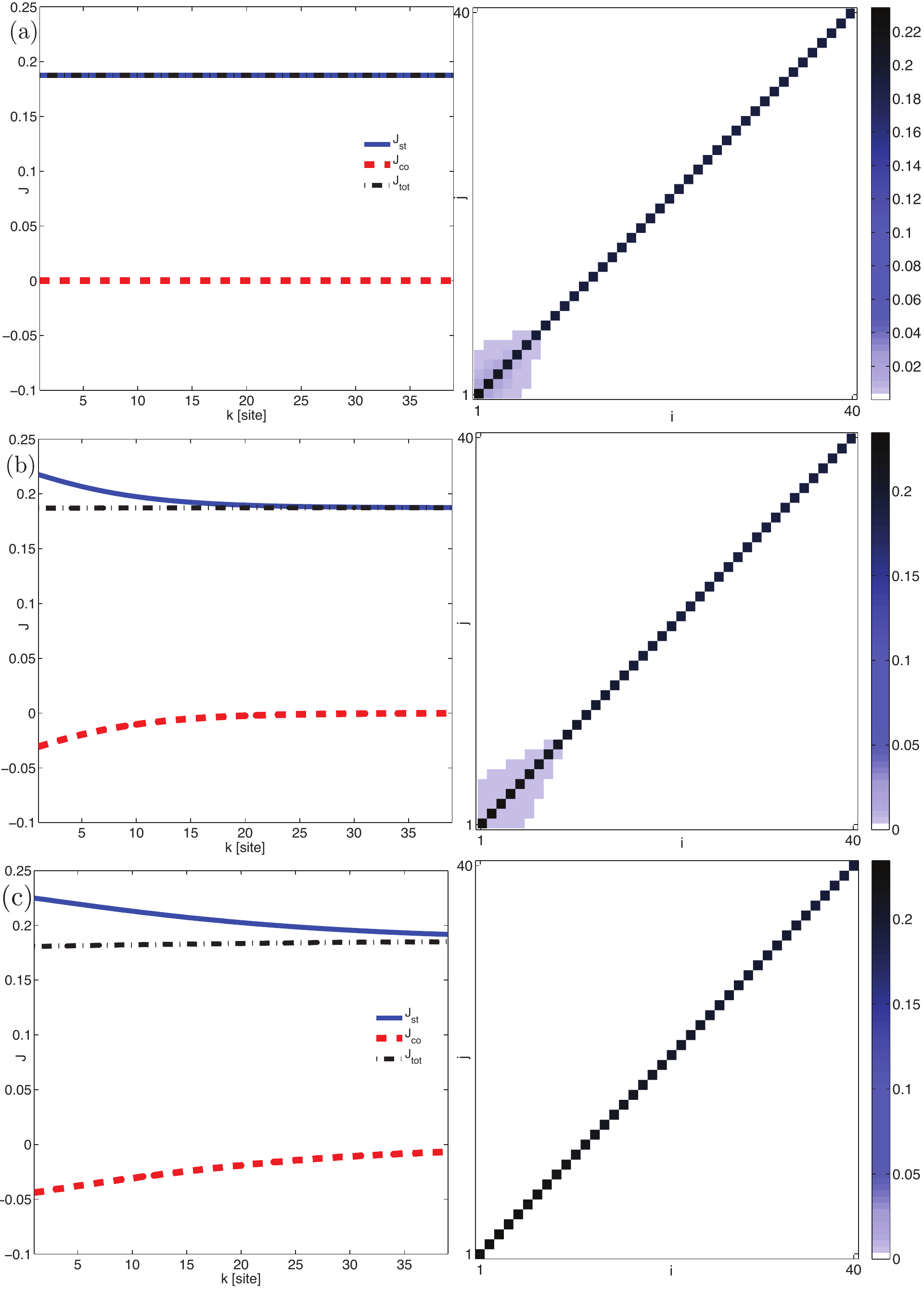}}
 \caption{\label{fig:HD_corr} (color online)
HD phase:
The left column shows the density-density correlations (\ref{corr_f})  for $\alpha = 1/2$ and $\beta = 1/4$  (HD) 
and different values of coherent driving $\lambda$. In the right column the current-density is plotted as a 
function of the lattice site $k$. The black dash-dotted line depicts the total current $j^{tot}$. 
The blue solid line corresponds to the stochastic contribution $j^{st}$
and the red dashed line shows the coherent contribution $j^{co}$.
The plots are ordered from top to bottom as (a) $\lambda=0$ (b)$\lambda =  0.5$ and (c)$\lambda =  1$.}
\end{figure}
\end{center}

\begin{center}
\begin{figure}[ht]
\resizebox{0.95\linewidth}{!}
{\includegraphics{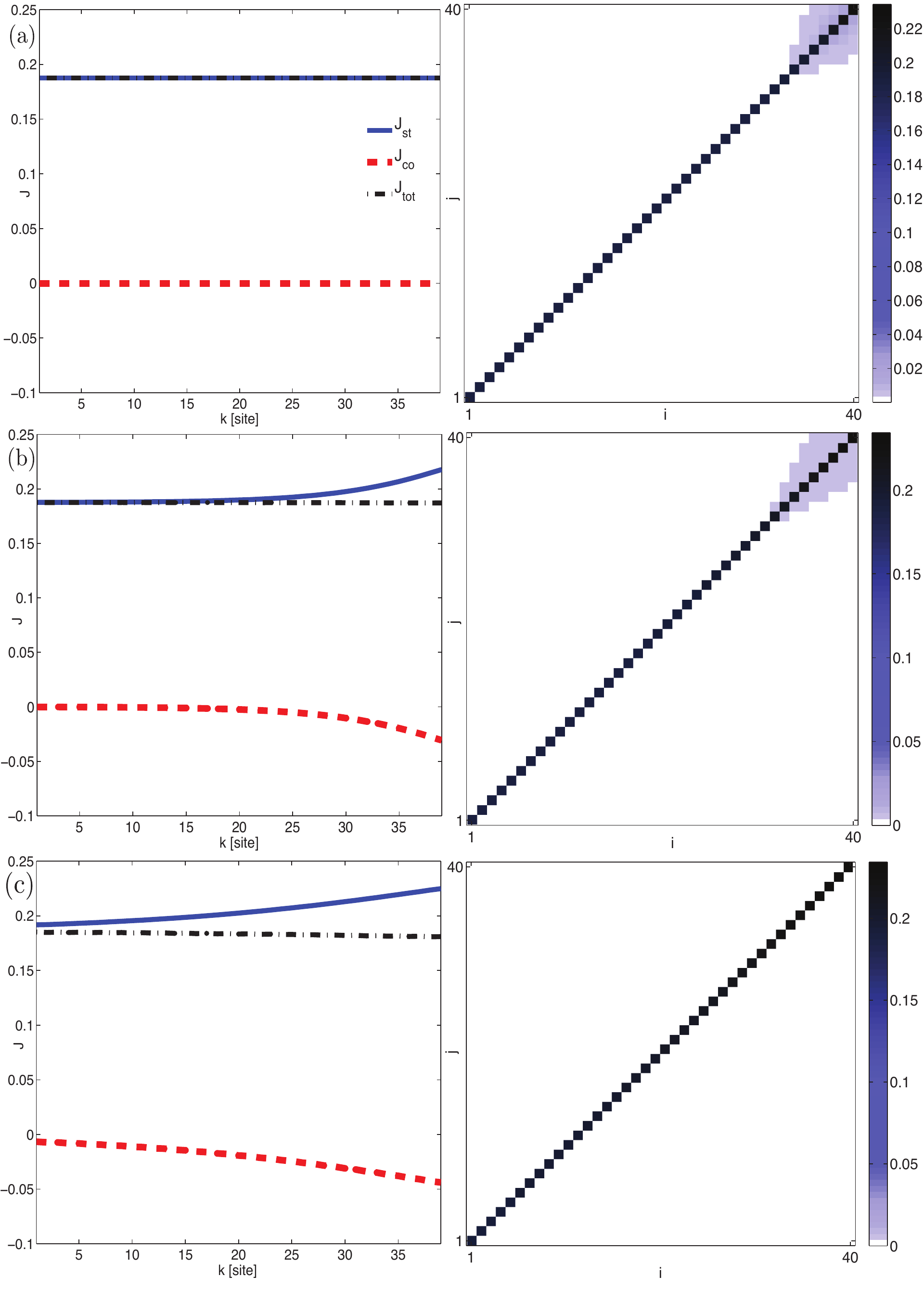}}
\caption{\label{fig:LD_corr} (color online)
LD phase:
 In the left column we plotted the density-density correlations (\ref{corr_f})  for $\alpha = 1/4$ and $\beta = 1/2$  (LD). 
 The right column depicts the current-density as a function of the lattice site $k$. The black dash-dotted line 
 amounts to the total current $j^{tot}$. The blue solid line corresponds to the stochastic contribution $j^{st}$
and the red dashed line shows the coherent contribution $j^{co}$.
The plots are ordered from top to bottom 
as (a) $\lambda=0$ (b)$\lambda =  0.5$ and (c)$\lambda =  1$.}
\end{figure}
\end{center}

\begin{center}
\begin{figure}[ht]
\resizebox{0.95\linewidth}{!}
{\includegraphics{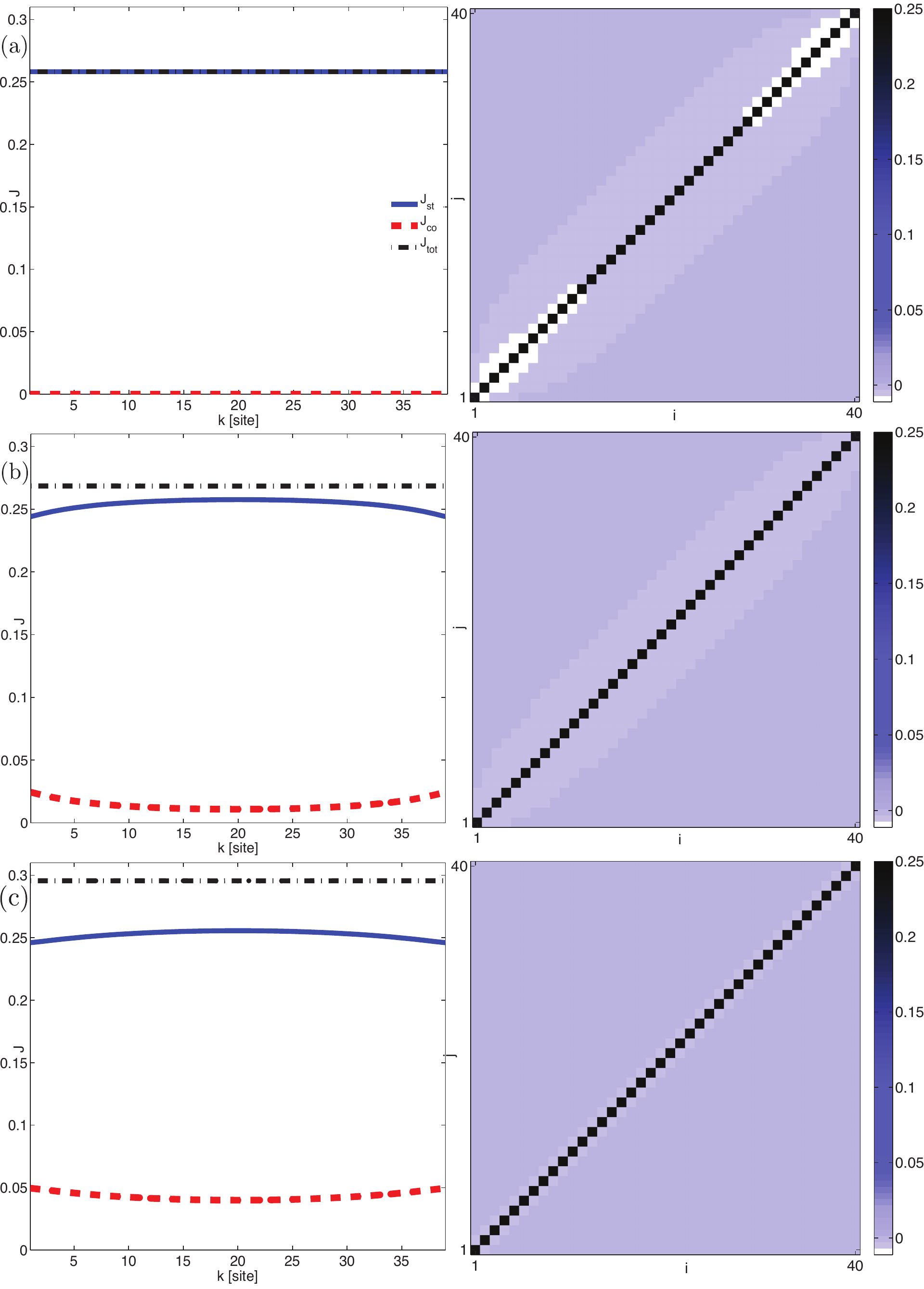}}
\caption{\label{fig:MC_corr} (color online)
MC phase:
The left column shows the density-density correlations (\ref{corr_f})  for $\alpha = 3/4$ and $\beta = 3/4$  (MC) 
and different values of coherent driving $\lambda$. In the right column the current-density is plotted as a 
function of the lattice site $k$. The black dash-dotted line depicts the total current $j^{tot}$. 
The blue solid line corresponds to the stochastic contribution $j^{st}$
and the red dashed line shows the coherent contribution $j^{co}$.
We plotted three different values for (a) $\lambda=0$ (b)$\lambda =  0.5$ and (c)$\lambda =  1$.}
\end{figure}
\end{center}

\begin{center}
\begin{figure}[ht]
\resizebox{0.95\linewidth}{!}
{\includegraphics{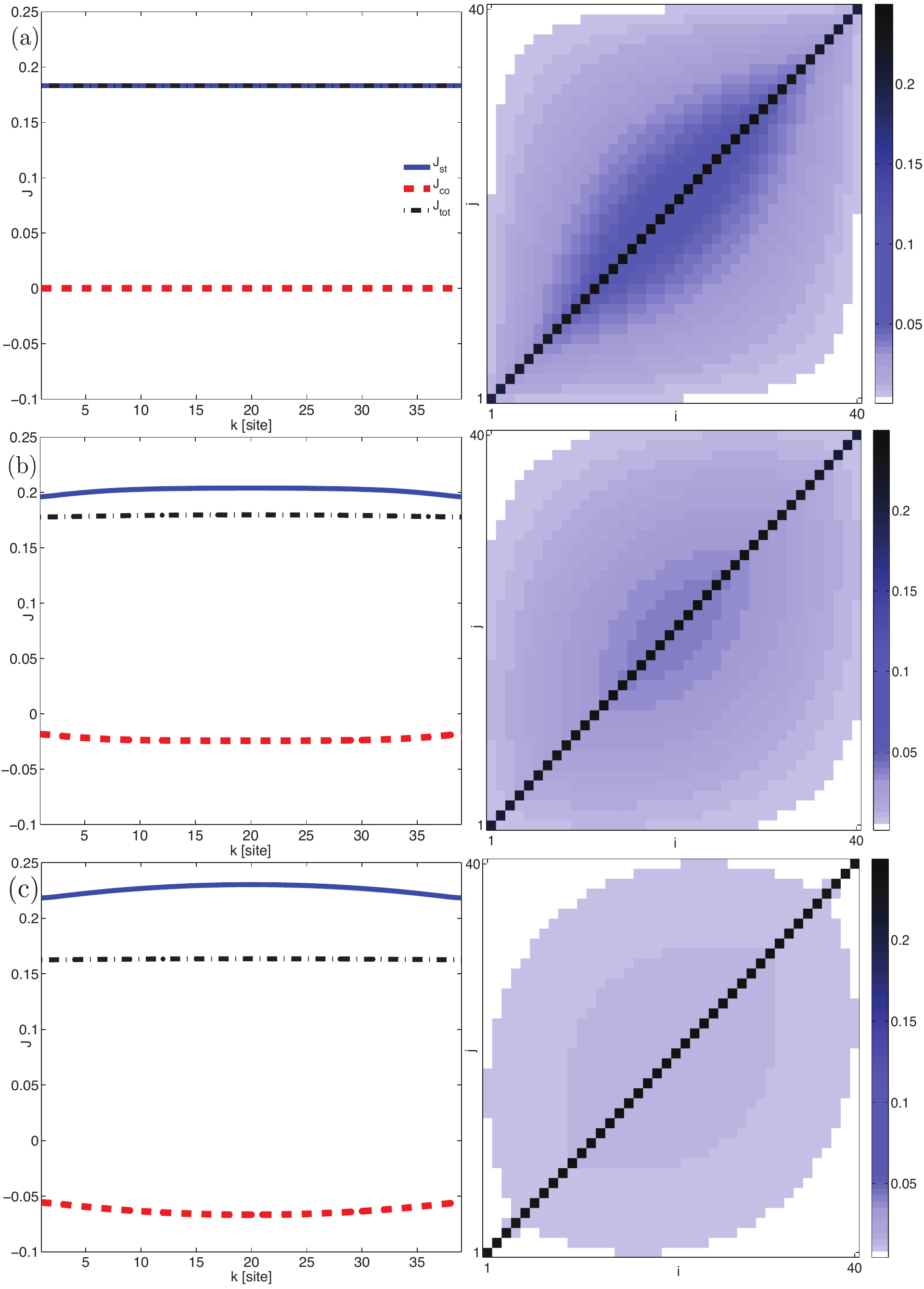}}
\caption{\label{fig:CL_corr} (color online)
Coexistence line:
The left column shows the density-density correlations (\ref{corr_f})  for $\alpha = 1/4$ and $\beta = 1/4$  (CL). 
In the right column the current-density is plotted. The black dash-dotted line amounts to the total current $j^{tot}$. 
The blue solid line corresponds to the stochastic contribution $j^{st}$ and the red dashed line shows the coherent 
contribution $j^{co}$. The values for the coherent driving are chosen as
(a) $\lambda=0$ (b)$\lambda =  0.5$ and (c)$\lambda =  1$.}
\end{figure}
\end{center}

The steady state of the ASEP, without any additional quantum evolution, i.e. $\lambda=0$, 
can be calculated exactly \cite{Derrida93} and its solution can be written in terms of  
a matrix product density operator (MPDO) \cite{fverstraete04}. The general form of a MPDO is given by,
\begin{equation}
\label{cl:sol1}
\rho = \sum_{i_1,\ldots , j_N=0}^{d-1} \bra{V}M_{i_1,j_1},\ldots , M_{i_N,j_N} \ket{W} \, \bigotimes_{k=1}^{N} \ket{i_k}\bra{j_k}, 
\end{equation}
where $d$ denotes the dimension of the local Hilbert space and $M_{i_k,j_k+1}$ are $D_k \times D_{k+1}$ 
dimensional matrices,   
In the case of the ASEP, the steady state solution of the master equation is given by the choice, see Appendix  \ref{app:steady}:

\begin{eqnarray}
\label{cl:sol2}
\bra{W} &=& \sum_{k=0}^{N} \frac{1}{\alpha^k}\bra{k} \\ \nonumber
M_{1,0} &=& M_{0,1} = 0 \\ \nonumber 
M_{0,0} &=& \sum_{k=0}^{N} \ket{k} \bra{k-1} \\ \nonumber
M_{1,1} &=& \sum_{n=0}^{N} \frac{1}{\beta} \ket{0}\bra{n} + \sum_{n=1}^{N}\sum_{m=1}^{n}\ket{m}\bra{n} \\ \nonumber
\ket{V} &=& \ket{0},
\end{eqnarray}

which is a MPDO with a matrix dimension of $D=N+1$.
In \cite{Derrida93} the classical phase diagram with respect to
$\alpha$ and $\beta$ has been found FIG. \ref{fig:phase_diagram}. 
In general there are three phases, the low-density phase LD, the high density phase HD, 
and the maximal current phase MC.
The three phases exhibit different behavior in the current and density. The statistical current and the 
corresponding density  in the large $N$ limit are given by,
\begin{center}
\begin{tabular}{r|c|c|c}
          &  LD                  &   HD   &   MC       \\\hline
$n$       & $\alpha $            &  $1-\beta $          &  $1/2$     \\
$j^{st}$  & $\alpha(1-\alpha)$   &  $\beta(1-\beta)$  &   $1/4$     \\
\end{tabular}
\end{center}
as has been calculated in \cite{Derrida93}. 
We want to understand the system's response to a quantum mechanical 
perturbation at distinct points in the phase diagram.\\
Note, there is a special line in the classical phase diagram marked by 
$\alpha+\beta=1$. Along this line the density operator for the classical probabilities factorizes, 
and mean-field theory becomes exact, Appendix \ref{app:steady}. 
If we now add the quantum perturbation,  we see, that the state,
\begin{equation}
\rho = \bigotimes_{k=1}^{N}\left(
\begin{array}{cc}
 \beta &    0     \\
   0   & \alpha 
\end{array}\right),
\end{equation} 
is still the steady state of the system for arbitrary values of $\lambda$. 
That is, when the system is classically uncorrelated, the quantum perturbation
has no effect on the system. \\
To see that in the regime, where the stochastic steady state is correlated, the coherent
evolution alters the steady state, we need to calculate the steady state of the system numerically
by time-evolving the density matrix, until we reach the steady state. 
The numerical simulations of the real time evolution is performed by a modification of the 
TEBD algorithm for mixed states \cite{fverstraete04,vidal04}, see Appendix \ref{app:numerics} 
for details. As initial state for the evolution we have chosen the classical steady state (\ref{cl:sol1}). 
We then changed the value of $\lambda = 0$ to, $\lambda = 1/2$ and $\lambda = 1$, and evolved the 
MPDO, until the steady state was reached, i.e. until all considered observables did not change any more.
Negative values of $\lambda$ were also simulated and led to the same results. We conclude from this,
that the systems response only depends on the absolute value of $\lambda$.  
The simulations were done for a lattice with $N=40$ sites. The matrix bond dimension of the MPDO was 
chosen as $D=60$ and we chose a Trotter step $\Delta t = 10^{-4}$. \\
To get an understanding of how the system responds to the quantum perturbation in each phase respectively,
we have chosen to compute the steady state at different values of $\alpha$ and $\beta$, which 
correspond to points chosen to lie in different phases. Four different points were chosen. Corresponding
to the coexistence line (CL), where $\alpha = \beta$ and $\beta+\alpha \leq 1$, we chose the point $\alpha = \beta = 1/4$.
In the maximum current phase (MC), we chose $\alpha = \beta = 3/4$. For the low-density (LD) and the high-density (HD) phase,
we chose $\alpha = 1/4 \;\; \beta = 1/2$ and $\alpha = 1/2 \;\; \beta= 1/4$ respectively.\\
The observable we considered first was the density distribution $\langle n_k \rangle = \Tr\left\{n_k \rho \right\}$ 
as a function of the lattice site $k$, FIG ~\ref{fig:dens}. 
Furthermore, we calculated the values of the two-point correlation functions, of the 
densities $n_k = \sigma^+_k\sigma^-_k$ for all pairs $(i,j)$ of sites,
\begin{equation}
\label{corr_f}
\langle n_i n_j \rangle^c = \langle n_i n_j \rangle - \langle n_i \rangle \langle n_j \rangle.
\end{equation}
The expectation values are taken with respect to the system's steady state.
In the figures FIG. ~\ref{fig:HD_corr},\ref{fig:LD_corr},\ref{fig:MC_corr},\ref{fig:CL_corr}, 
the correlation functions are compared to the different contributions to the 
current $j^{tot}$ defined in (\ref{curr}), for different values of $\lambda=0,1/2,1$.
\vspace{-2.3cm}
The first observation to be made is that the individual contributions to the total current 
are no longer constant throughout the lattice anymore, they show a dependence on the lattice site. 
The total current however, i.e. $j_{tot} = j_{co} + j_{st}$, is still constant at each site of the lattice,
as is required, since the system is in a steady state.
\\

\paragraph{The high- and low-density phases FIG.~\ref{fig:HD_corr},\ref{fig:LD_corr}:}
These two phases are, just as in the classical set up, related by a particle-hole transformation 
and exchanging the numbering of the lattice sites from left to right. All the plots reflect this 
symmetry. One observes that the correlation functions in the classical regime, $\lambda = 0$, are 
already quite short ranged and decay rapidly. The quantum perturbation in both  cases leads to a 
further decay of the correlations. The total current remains stable with respect to the quantum 
perturbation and does not change its value notably. The individual constituents to the current
however change their behavior. At the boundaries the stochastic 
contributions are increased, whereas the coherent current gives rise to a flow in the opposite 
direction.     

\paragraph{The maximal current phase FIG.~\ref{fig:MC_corr}:}
In the classical process, the MC phase corresponds to the maximum amount of current the system can carry. 
Allowing for a quantum perturbation, the system makes use of the additional transport capacity and increases 
its total current. For these boundary conditions the stochastic as well as the coherent contributions flow
in the same direction. Note, however, that the system in the bulk prefers to use the stochastic transport,
whereas at the boundaries coherent transport is preferred. The classical correlation function initially 
assumes negative values close to the boundaries. The onset of the quantum perturbation also reduces the 
magnitude of the correlations in this phase, even though the total amount of current is increased.   

\paragraph{Coexistence line FIG.~\ref{fig:CL_corr}:}
For the chosen boundary conditions, the amount of correlations initially present in the steady-state are 
decreased, when switching on the quantum perturbation. The final steady-state, were $\lambda=1$, however, 
still shows the presence of correlation to a higher degree than in the other phases. The total amount
of current carried by the system, however, is decreased. The coherent contribution is negative throughout the 
system. 
\section{The Symmetric exclusion process}
\label{sec:SEP}
\begin{center}
\begin{figure}[ht]
\resizebox{\linewidth}{!}
{\includegraphics{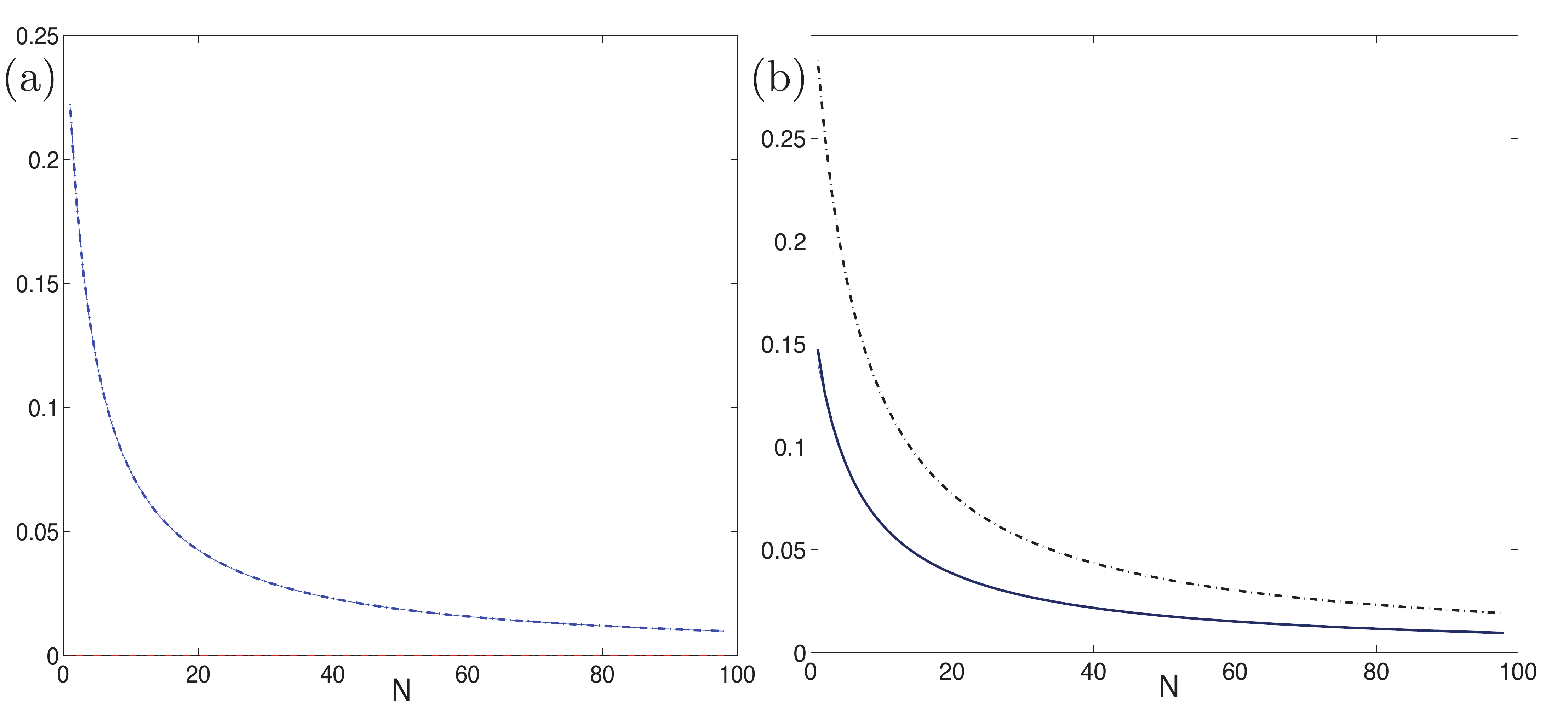}}
\caption{\label{fig:SEP_scale}(color online) The plots demonstrate the scaling 
of the currents for large  $N$. Plot (a) depicts the scaling of the total current $j^{tot} = j^{st}$,blue dashed line  
for $\lambda = 0$. The plot (b) corresponds to $\lambda = 1$. Here the total current $j^{tot}$ is represented by the 
black dash-dotted line. The coherent contribution corresponds to the blue solid line. 
The boundary values were chosen as $\alpha = \beta = 3/4$}
\end{figure}
\end{center}

\begin{center}
\begin{figure}[ht]
\resizebox{\linewidth}{!}
{\includegraphics{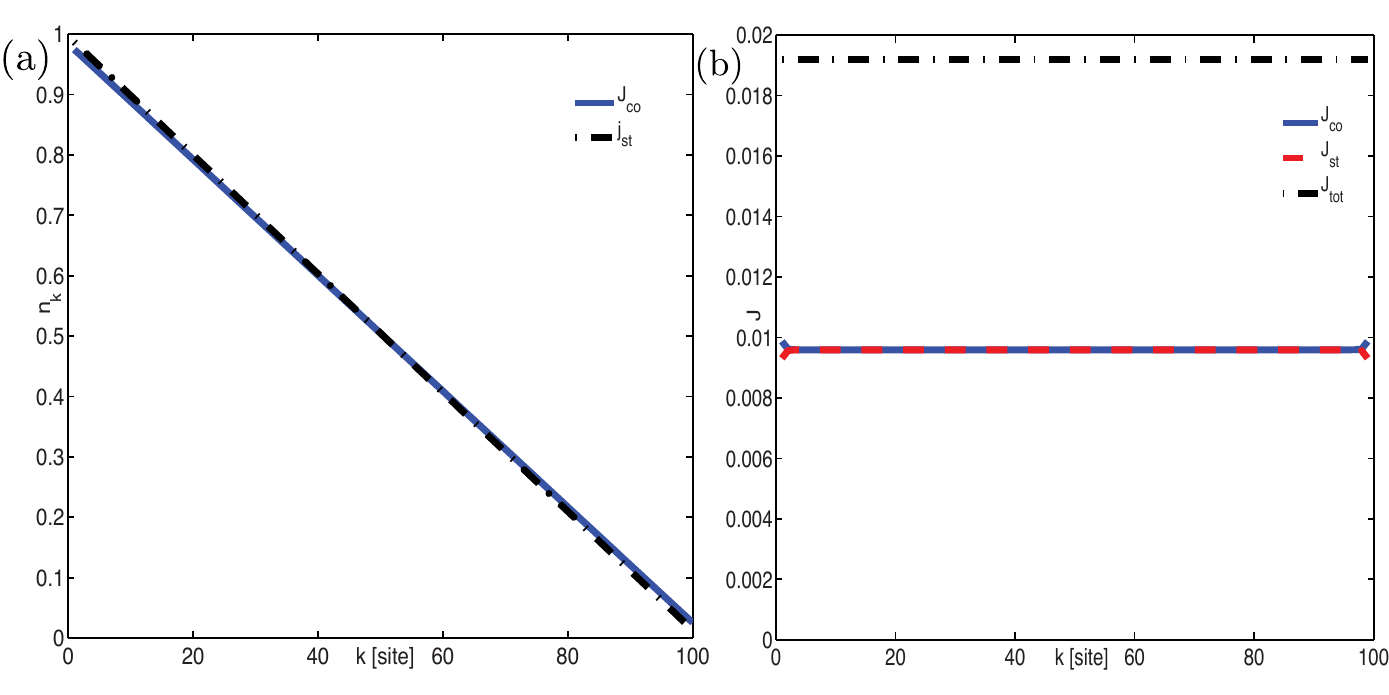}}
\caption{\label{fig:SEP_Jdens} (color online) 
 Plot (a) shows the density distributions for the values $\lambda = 0,1$ and for $\alpha = \beta  = 3/4$ .
 Calculations for larger system sizes indicate, that this deviation vanishes in the thermodynamic limit. 
 The second plot (b)  shows the dependence of the current distribution on the lattice site for $\lambda = 1$.  
 The stochastic contribution $j_{st}$ as well as the coherent contribution $j_{co}$ only marginally differ at the boundaries}
\end{figure}
\end{center}

If we allow for stochastic hopping in both directions with an equal rate $\varphi$ and turn off the coherent evolution, 
the model describes the classical symmetric exclusion process.
The symmetric exclusion process is known to possess only a single phase classically with a vanishing current \cite{SEP1}.
We now consider the quantum perturbation to the system. Note, that now the quantum-jump operators are related via 
$L^{R}_{k,k+1} = {L^{L}}^{\dagger}_{k,k+1} \equiv L_{k,k+1} $. This allows us to rewrite the full master equation as,
\begin{eqnarray}
\label{master_sep}
\partial_t \rho &=& -i\left [\rho ; H \right] \\ \nonumber 
 &+&\sum_{k=1}^{N-1}[[L_{k,k+1};\rho]; L_{k,k+1}^{\dagger}]
 +[[L_{k,k+1}^{\dagger};\rho]; L_{k,k+1}] \\ \nonumber  
 &+& L_1\rho L_1^{\dagger} -\frac{1}{2}\{L_1^{\dagger},L_1;\rho\}
 + L_N\rho L_N^{\dagger} -\frac{1}{2}\{L_N^{\dagger},L_N;\rho\}.
\end{eqnarray}  
It is possible to calculate the nearest neighbor two-point correlation functions exactly. To see why this is possible, 
we first transform the Pauli raising and lowering operators, $\sigma^+$ and $\sigma^-$ to fermionic modes by means of 
the Jordan-Wigner transformation. The fermionic modes read then,
\begin{eqnarray}
a_k^{\dagger} = -\left( \bigotimes_{i=1}^{k-1}\sigma^{z}\right)\, \sigma^{+} \\ \nonumber 
a_k                 = -\left( \bigotimes_{i=1}^{k-1}\sigma^{z}\right)\, \sigma^{-}
\end{eqnarray}
One can verify, that these modes now obey the fermionic anti-commutation relations, $\{a_k;a_l^{\dagger}\} = \delta_{k,l}$.
It is possible to calculate the evolution of the fermionic two-point function $\la a_k^{\dagger}a_m \ra$ from the master 
equation (\ref{master_sep}) via, 
\begin{eqnarray}
\partial_t \la a_k^{\dagger}a_m \ra = \Tr\left\{{\cal L}^{\dagger}[a_k^{\dagger}a_m] \rho \right \} 
\equiv \la {\cal L}^{\dagger}[a_k^{\dagger}a_m] \ra.
\end{eqnarray}  
Since the commutator of two  pairs  of fermionic modes is again an operator made up from two fermionic modes, we see, that the time-evolution 
of the fermionic two-point functions again only depends on two-point functions. So the two-point correlation functions of the steady 
state can be computed exactly. The equations for the correlation functions read,
\begin{eqnarray}
\label{eq:diag}
&&\partial_t \la a^{\dagger}_m a_l \ra = 							                     \no
&&\frac{\alpha}{2}(\delta_{m,0} + \delta_{l,0}) \la a_l a^{\dagger}_m \ra -
\frac{\beta}{2} (\delta_{m,N} + \delta_{l,N}) \la a^{\dagger}_m a_l \ra                              \no
&&-\frac{\varphi}{2}\left([\la a_{m+1}a_{m+1}^{\dagger}\ra - \la a_{m+1}^{\dagger}a_{m+1}\ra \right. \no
&&+ \la a_{m-1}a_{m-1}^{\dagger}\ra - \la a_{m-1}^{\dagger}a_{m-1}\ra ]\delta_{l,m}                  \no
&&+ \left. 2[\la a_{m}^{\dagger}a_{l}\ra -\la a_{l}a_{m}^{\dagger}\ra] \right )                      \no
&&- i\lambda \left([\la a_{m}^{\dagger}a_{l+1}\ra + \la a_{m}^{\dagger}a_{l}-1\ra] \right.           \no
&&- \left.[\la a_{m+1}^{\dagger}a_{l}\ra +\la a_{m-1}^{\dagger}a_{l}\ra] \right ),
\end{eqnarray}
and,
\begin{eqnarray}
\label{eq:offdg}
&&\partial_t \la a^{\dagger}_m a^{\dagger}_l \ra = 									     \no
&&-\frac{\alpha}{2}\left(\delta_{0,m} + \delta_{0,l}\right) \la a^{\dagger}_m a^{\dagger}_l \ra 
-\frac{\beta}{2}\left(\delta_{l,N}+\delta_{m,N}\right)\la a^{\dagger}_m a^{\dagger}_l \ra                  \no
&&-\varphi\left(\left[\delta_{l+1,m}\la a^{\dagger}_l a^{\dagger}_{l+1} \ra  \right.\right. 
-\left.\left.\delta_{m+1,l}\la a^{\dagger}_m a^{\dagger}_{m+1} \ra \right] \right.                      
+ \left. 2\la a^{\dagger}_m a^{\dagger}_l \ra \right)                                                                           \no
&&+i\lambda\left(\la a^{\dagger}_m a^{\dagger}_{l-1} \ra + \la a^{\dagger}_m a^{\dagger}_{l+1} \ra \right.
 \left. +\la a^{\dagger}_{m-1} a^{\dagger}_l \ra + \la a^{\dagger}_{m+1} a^{\dagger}_l \ra \right).      
\end{eqnarray}
The other correlation functions are related to these two by the identities imposed due to the anti-commutation 
relations of the fermionic modes, thus $\la a^{\dagger}_m a^{\dagger}_l\ra = \la a_m a_l\ra^{*}$ and
$\la a^{\dagger}_m a_l\ra = \delta_{l,m} - \la a_l a_m{\dagger} \ra^{*}$.\\
The steady-state correlations can be computed from these equations by requiring stationarity, i.e.
$\partial_t \la a^{\dagger}_m a_l\ra = \partial_t \la a^{\dagger}_m a_l^{\dagger}\ra = 0$. One needs 
to solve the corresponding system of linear difference equations.\\
The current as well as the particle number density can be expressed in terms of these correlators. 
One finds for the particle number density $\la n_k \ra = \la a_k^{\dagger}a_k \ra$ and the two contributions to the current read,
\begin{eqnarray}
&j^{st}_{k,k+1} = \varphi\left ( \la n_{k} \ra - \la n_{k+1} \ra \right ) \\ \nonumber
&j^{co}_{k,k+1} = \frac{2\lambda}{i}\left(\la a_{k}^{\dagger}a_{k+1} \ra - \la a_{k} a_{k+1}^{\dagger} \ra \right). 
\end{eqnarray} 
Note, that the stochastic current now only depends on the difference of the densities at adjacent sites.  
The equations (\ref{eq:diag}) and (\ref{eq:offdg}) are solved numerically for different parameters 
$\alpha$ and $\beta$.\\
The classical SEP without any further driving obeys the detailed balance condition, and thus 
does not  support a steady state current. When one allows for an external driving of the particles at the 
boundaries, as we do in our example, a current is induced in the SEP steady state. This current however vanishes 
as $\sim 1/N$ in the system size $N$, see FIG. \ref{fig:SEP_scale}(a) .  If  we turn to the coherent part,
we also see this behavior, i.e. the current vanishes in the limit $N \rightarrow \infty$ FIG. \ref{fig:SEP_scale}(b).
We deduce from this, that the quantum perturbation to the SEP is an irrelevant perturbation and does not lead 
to a qualitatively different behavior of the systems transport properties. This point is also affirmed by considering
the particle density distribution as seen in FIG. \ref{fig:SEP_Jdens}(a). Here one sees that the density distribution 
in the presence of a quantum perturbation almost coincides with the classical distribution. The different contributions
to the current, FIG. \ref{fig:SEP_Jdens}(b), are identical in the system's bulk, and add up to an increased current
$J_{tot}$. Whether the quantum perturbation is completely irrelevant however can not be deduced from just 
considering the steady state density and the current alone. Here one would need to take higher order 
correlations into account, as for instance the current-current correlation function at unequal times.   

\section{Conclusions and outlook}
\label{sec:Concl}
We have investigated a quantum perturbation to the dynamics of the stochastic asymetric exclusion process
as well as to the symmetric exclusion process. We find that we can rephrase the  stochastic master 
equation as a quantum equation that fully reproduces the classical dynamics. The quantum perturbations 
modify the steady state behavior and allow for two different types of currents, which  each on their own
can vary as a function of the site. Numerical simulations of the full master equation indicate, that the 
underlying classical phase-diagram of the stochastic process is respected. The steady state 
responds to driving due to the boundary terms with a different behavior in current and 
density. A further step would be to investigate the current-current correlation 
function of the SEP, to see whether the quantum perturbation has an effect to the current-fluctuations.
Furthermore,other, more complex models with an interplay between stochastic and coherent dynamics can
be investigated along these lines. 
\section{Acknowledgments}
\noindent
This work was supported by the FWF doctoral program Complex Quantum Systems (W1210).

\begin{appendix}
 
\section{Derivation of the steady state}
\label{app:steady}
\noindent
The steady state can be calculated in matrix product form by applying the 
Derrida method of the quadratic algebra \cite{Derrida93}. To better understand this, we 
shall rewrite the MPDO with translationally invariant matrices $M_{i,j}$ as,
\begin{eqnarray}
\rho = \bra{W}\left(\begin{array}{cc} M_{00} & M_{01} \\ M_{10} & M_{11}\end{array}\right)^{\otimes N}\ket{V} 
\end{eqnarray}
This MPDO now serves as steady state  ansatz for the master equation. To derive the quadratic algebra, consider
the total master equation (\ref{master}), where we now regrouped the terms in local two-body interactions and the single 
particle boundary terms. This can be written as,  
\begin{eqnarray}
\label{sumSt}
{\cal L}[\rho] = {\cal L}_1[\rho] +  \sum_{k=1}^{N-1}{\cal L}_{k,k+1}[\rho] + {\cal L}_1[\rho].
\end{eqnarray}
We now try to find relations for the matrices $M_{i,j}$ so that $\rho$ will correspond to the steady state 
of the system. We introduce additional ancilla matrices $ \hat{M}_{ij}$ and require, that
\begin{eqnarray}
{\cal L}_{k,k+1}\left[\left(\begin{array}{cc} M_{0,0} & M_{0,1}\\ M_{1,0} & M_{1,1}\end{array}\right)\right.
\left. \otimes \left (\begin{array}{cc} M_{0,0} & M_{0,1} \\ M_{1,0} & M_{1,1} \end{array} \right ) \right] = \no
\left (\begin{array}{cc} M_{0,0} & M_{0,1} \\ M_{1,0} & M_{1,1} \end{array} \right ) \otimes
\left (\begin{array}{cc} \hat{M}_{0,0} & \hat{M}_{0,1} \\ \hat{M}_{1,0} &\hat{M}_{1,1} \end{array} \right ) - \no 
\left (\begin{array}{cc} \hat{M}_{0,0} & \hat{M}_{0,1} \\ \hat{M}_{1,0} &\hat{M}_{1,1} \end{array} \right ) \otimes
\left (\begin{array}{cc} M_{0,0} & M_{0,1} \\ M_{1,0} & M_{1,1} \end{array} \right ).
\end{eqnarray}
Furthermore we require, that the single site operators at the boundaries have to satisfy,
\begin{eqnarray}
 \bra{V}{\cal L}_1\left[\left (\begin{array}{cc} M_{0,0} & M_{0,1} \\ M_{1,0} & M_{1,1} \end{array} \right )\right]  = 
 \bra{V}\left (\begin{array}{cc} \hat{M}_{0,0} & \hat{M}_{0,1} \\ \hat{M}_{1,0} & \hat{M}_{1,1} \end{array} \right )   \no
{\cal L}_N\left[\left (\begin{array}{cc} M_{0,0} & M_{0,1} \\ M_{1,0} & M_{1,1} \end{array} \right )\right]\ket{W} = 
 -\left (\begin{array}{cc} \hat{M}_{0,0} & \hat{M}_{0,1} \\ \hat{M}_{1,0} & \hat{M}_{1,1} \end{array}  \right )\ket{W}.                            
\end{eqnarray}
We see that the total sum (\ref{sumSt}) telescopes to zero and $\rho$ is the steady state solution of the 
equation. To find an explicit  solution one needs to construct a representation of the matrices $M_{ij}$ that 
obey the given algebraic constraints. In the simple case, where $\lambda=0$ and $\varphi = 1$, 
we reproduce the known classical algebra for the ASEP steady state \cite{Derrida93}. Here one chooses 
$M_{0,1} = M_{1,0} = 0$, as well as $ \hat{M}_{0,1} = \hat{M}_{1,0} = 0$. The diagonal terms, upon 
stetting, $\hat{M}_{0,0} = -\hat{M}_{1,1} = 1$, have to satisfy the algebra $M_{1,1}M_{0,0} = M_{0,0} + M_{1,1}$.
The boundary terms have to fulfill, $\bra{V}M_{0,0} = 1/{\alpha}\bra{V}$ and $M_{1,1}\ket{W} = 1/{\beta}\ket{W}$. 
One possible representation for this algebra is given  by (\ref{cl:sol2}), when considering $N \rightarrow \infty$. 
Note, that when one wants to reproduce the steady state of a finite system of size $N$, it suffices to 
choose a matrix dimension of $D=N+1$ for this specific representation.  
One can also verify, that at  $\alpha+\beta=1$ the representation of the algebra can be 
chosen one-dimensional \cite{Derrida93}. 
\section{Numerical Implementation}
\label{app:numerics}
For the numerical evolution of the density-matrix according to (\ref{master}), we apply a numerical scheme developed 
in \cite{fverstraete04}. Starting from the initial density-matrix $\rho_0$ given as an MPDO, we apply the CP-map 
${\cal E}(t) = \exp(t {\cal L})$ for a small time step $\Delta t$ and approximate the resulting density operator,
that has now an increased bond dimension, with an MPDO that has a bond-dimension $D_k$ corresponding to that of the
original MPDO. The approximation of the operator $\rho(t+\Delta t) = {\cal E}(\Delta t)\, \rho(t)$ is chosen, such 
that the Hilbert-Schmidt norm 
$\|\rho(t+\Delta t) - \rho_{new} \|^2_{HS} = \Tr \left[\left(\rho(t+\Delta t) - \rho_{new}\right)^2 \right]$ 
is minimized.This optimization can be performed efficiently by sweeping from left to right over the individual sites and 
optimizing the matrices locally. For the application of the CP-map to be computable, we
perform a second-order Trotter expansion of the CP-map as follows,
\begin{eqnarray}
\label{trotter}
{\cal E}({\cal L},\Delta t) \simeq {\cal E}({\cal L}_o,\Delta t/2){\cal E}({\cal L}_e,\Delta t){\cal E}({\cal L}_o,\Delta t/2),
\end{eqnarray}
where ${\cal L} = {\cal L}_e+{\cal L}_o$  corresponds to splitting the Liouvillian into commuting terms which act on the
sites $(2k,2k+1)$ and $(2k-1,2k)$, respectively. The resulting MPDO $\rho_{new}$ is then chosen as initial condition for 
the next step and the procedure is repeated.   
\end{appendix}

\end{document}